\documentclass[twocolumn]{aastex631}
\usepackage{tikz}
\usepackage{subfigure}
\usepackage{threeparttable}
\usepackage{appendix}
\usetikzlibrary{positioning, shapes.geometric}
\usepackage{multirow}

\usepackage{hyperref}

\shorttitle{A Direct Method to Compute Doppler Beaming Factors}
\shortauthors{Zheng et al.}

\begin{document}
\title{Direct Method to Compute Doppler Beaming Factors in Binary Stars}
\author{Chuanjie Zheng}
\affiliation{National Astronomical Observatories, Chinese Academy of Sciences, Beijing 100101, People's Republic of China;}
\affiliation{School of Astronomy and Space Science, University of Chinese Academy of Sciences, Beijing 100049,  People's Republic of China;}
\author{Yang Huang}
\affiliation{School of Astronomy and Space Science, University of Chinese Academy of Sciences, Beijing 100049,  People's Republic of China;} 
\affiliation{National Astronomical Observatories, Chinese Academy of Sciences, Beijing 100101, People's Republic of China;}
\author{Jifeng Liu}
\affiliation{National Astronomical Observatories, Chinese Academy of Sciences, Beijing 100101, People's Republic of China;}
\affiliation{School of Astronomy and Space Science, University of Chinese Academy of Sciences, Beijing 100049,  People's Republic of China;}
\affiliation{New Cornerstone Science Laboratory, National Astronomical Observatories, Chinese Academy of Sciences, Beijing 100101, People’s Republic of China;}
\affiliation{Institute for Frontiers in Astronomy and Astrophysics, Beijing Normal University, Beijing, 102206, People's Republic of China}
\author{Youjun Lu}
\affiliation{National Astronomical Observatories, Chinese Academy of Sciences, Beijing 100101, People's Republic of China;}
\affiliation{School of Astronomy and Space Science, University of Chinese Academy of Sciences, Beijing 100049,  People's Republic of China;} 
\author{Henggeng Han}
\affiliation{National Astronomical Observatories, Chinese Academy of Sciences, Beijing 100101, People's Republic of China;}
\author{Yuan Tan}
\affiliation{National Astronomical Observatories, Chinese Academy of Sciences, Beijing 100101, People's Republic of China;}
\affiliation{School of Astronomy and Space Science, University of Chinese Academy of Sciences, Beijing 100049,  People's Republic of China;}
\author{Timothy C. Beers}
\affiliation{Department of Physics and Astronomy and JINA Center for the Evolution of the Elements (JINA-CEE), University of Notre Dame, Notre Dame, IN 46556, USA}

\correspondingauthor{Yang Huang}

\begin{abstract}
The Doppler beaming effect, induced by the reflex motion of stars, introduces flux modulations and serves as an efficient method to photometrically determine mass functions for a large number of close binary systems, particularly those involving compact objects.
In order to convert observed beaming-flux variations into a radial-velocity curve, precise determination of the beaming factor is essential. Previously, this factor was calculated as a constant, assuming a power-law profile for stellar spectra. In this study, we present a novel approach to directly compute this factor. Our new method not only simplifies the computation, especially for blue bands and cool stars, but also enables us to evaluate whether the relationship between beaming flux and radial velocity can be accurately described as linear.
We develop a python code and compute a comprehensive beaming-factor table for commonly used filter systems covering main-sequence, subgiant, and giant stars, as well as hot subdwarf and white dwarf stars.
Both the code and our table are archived and publicly available at \href{https://doi.org/10.5281/zenodo.13049419}{Zenodo}.
\end{abstract}
\keywords{Binary stars (154), Radial velocity (1332), Computational methods (1965)}
\section{Introduction}
\label{sec:intro}

Doppler beaming refers to the apparent brightness variations caused by reflex motions of stars in binary systems. 
In cases where the velocities of targets are far from the speed of light, this effect can be approximately described by a linear relation between the target's radial movement and the observed flux \citep{BEAM_planet,BEAM_AMP,claret}. 
Thus, this effect can be utilized to derive radial-velocity curves for binary systems photometrically, which are otherwise traditionally measured through expensive and time-consuming spectroscopic observations.
However, the modulations in light curves are too small to be detected by ground-based telescopes.
Even considering close binary system with radial-velocity semi-amplitude $K$ from a few tens to hundreds of km\,s$^{-1}$, the photometric variations are only on the order of $K/c \sim 10^{-4} - 10^{-3}$, where $c$ is the velocity of light.

Thanks to great advances in space-based missions, including CoRoT \citep{COROT,corot2}, Kepler \citep{kepler}, and TESS \citep{TESS}, continuous light curves with ultra-high precisions of 50-100 parts per million (ppm) are now available for a large sample of field stars.
Doppler beaming effects have been clearly detected for binary systems from these precise light curves \citep{BEAM_Detect_space,BEam_obs_1,BEER0}.
The largest published sample, 70 beaming binaries, has been described by \citet{BEER2}, who applied a dedicated search algorithm for the light curves obtained by CoRoT.
The sample size of such binary systems will increase steadily through the continuously released data from the ongoing TESS survey.
Moreover, we can expect to discover distant beaming binaries even at the Galactic center using data from the upcoming China Space Station Telescope (CSST) bulge survey.

It is non-trivial to extract physical parameters of binary systems from the beaming effects by precise light curves alone.
The Doppler-beaming factor, characterizing the flux variation caused by the radial motion of star in a binary, is also required to be precisely determined. 
In the field of high-energy astrophysics, this factor is usually expressed as $\delta^{3-\alpha}$, by assuming the emission spectrum to following a power-law profile with an index of $\alpha$ (i.e., $F\propto\nu^{\alpha}$), where $\delta$ is the Doppler factor\footnote{$\delta$ is defined as $\frac{\sqrt{1-(v/c)^2}}{1+v_r/c}$.} \citep{3-a,3-a_H}.
On top of that, a constant monochromatic Doppler beaming factor of $3-\alpha$, characterising the linear relation between observed flux and radial velocity, is derived by \cite{BEAM_planet} for binary or planet plus host-star systems, given an orbital speed much smaller than $c$ in the Doppler factor. 
For a given filter, the photon-weighted bandpass-integrated beaming factor can be derived by convolving the filter's transmission curve with the synthetic spectrum of different types of star \citep[e.g.,][]{BEam_obs_1}.
In this way, a systematic grid of beaming factors, for dozens of commonly used filters (such as in the SDSS, $Kepler$, TESS, and $Gaia$ photometric systems), were calculated for white dwarfs, 
main-sequence, and giant stars by \cite{claret}.

However, all of these calculations are based on the assumption that stellar spectra can be 
well-represented by a power-law profile. 
In reality, stellar spectra often deviate significantly from this profile. For instance, the blue range of optical spectra is heavily absorbed due to the blanketing effects of metallic lines. Additionally, the entire optical spectra of late-type stars are dominated by molecular bands. 
These factors may lead to deviations from a constant beaming factor, or at least add complexities to its calculation.
To overcome these limitations, we compute the beaming flux at various radial velocities, instead of calculating a constant factor based on the linear relation derived by assuming a power-law stellar spectrum.
Using this novel method, we can first determine whether a constant beaming factor adequately describes the relation between beaming flux and radial velocity. 
If it does indeed exhibit linearity, as assumed in previous studies, this factor can be calculated more straightforwardly using this new approach.

This paper is organised as follows. 
Section\,2 describes our method in detail. 
Section\,3 presents our main results and compares them to previous studies.
Finally, a summary is given in Section\,4.

\begin{figure*}
    \centering
    \includegraphics[scale=0.65]{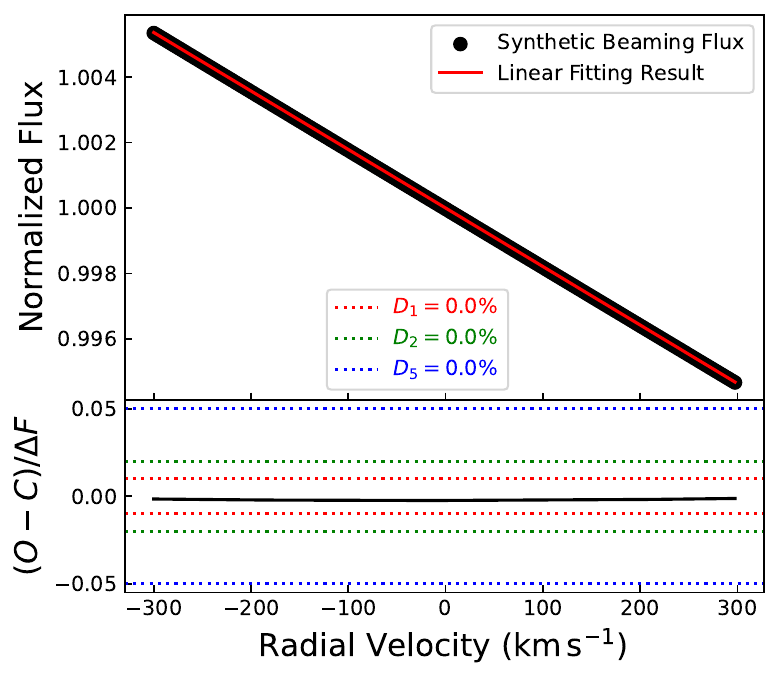}
    \includegraphics[scale=0.65]{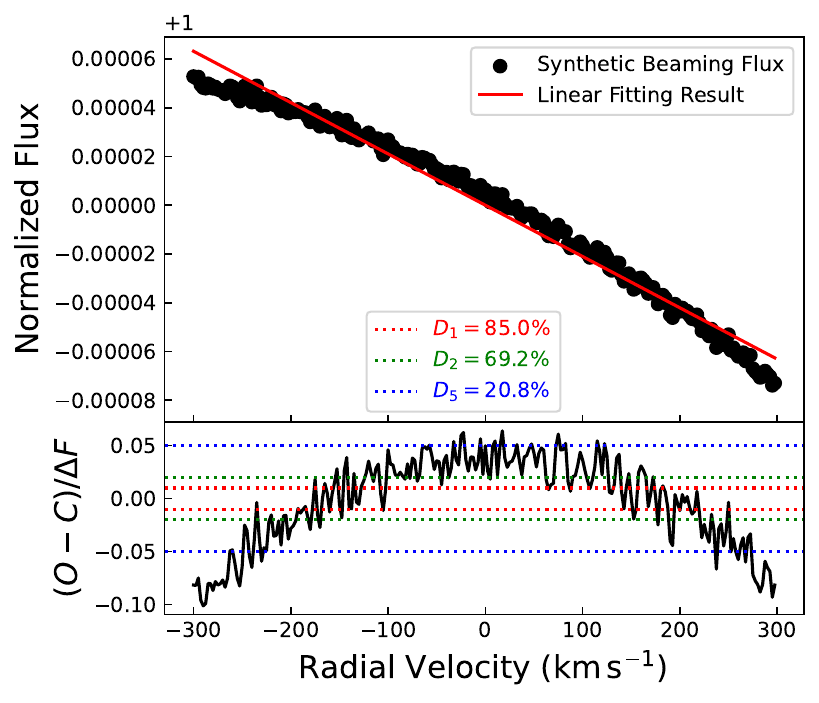}    
    \caption{Synthetic {\it B}-band beaming flux at different radial velocities for two binary systems with circular orbits ($e = 0$) and semi-amplitude radial velocity $K = 300$\,km\,s$^{-1}$. 
    In the left panel, the primary is an extremely metal-poor Solar-type star with $T_{\rm{eff}}$ = 5600\,K, log\,$g$ = 4.0, [Fe/H] = $-$3.0, and $E(B-V) = 0.0$.
    In the right panel, the primary is a metal-rich cool giant star with $T_{\rm{eff}} = 2600$\,K, log\,$g$ = 2.0, [Fe/H] = $+0.5$, and moderate reddening, $E(B-V) = 0.3$. 
    The red solid lines in both upper panels represent the best linear fit to the synthetic-beaming flux at the $B$-band, as a function of radial velocity variation of the binary system. The bottom panels show the relative-fitting residuals, defined as the ratio between the fit difference $O - C$ and the maximum flux variation $\Delta F$ (see text for definition) of the linear fits.
    The red, green, and blue dotted lines in the lower panels mark ratios of $\pm 1\%$, $\pm 2\%$ and $\pm5\%$, respectively. The corresponding deviation values of the $D_1$, $D_2$, and $D_5$ indexes (see text for definition) are labelled at the bottom of each upper panel.}
\label{fig:applicable}
\end{figure*}

\begin{figure*}
    \centering
    \includegraphics[scale=0.4]{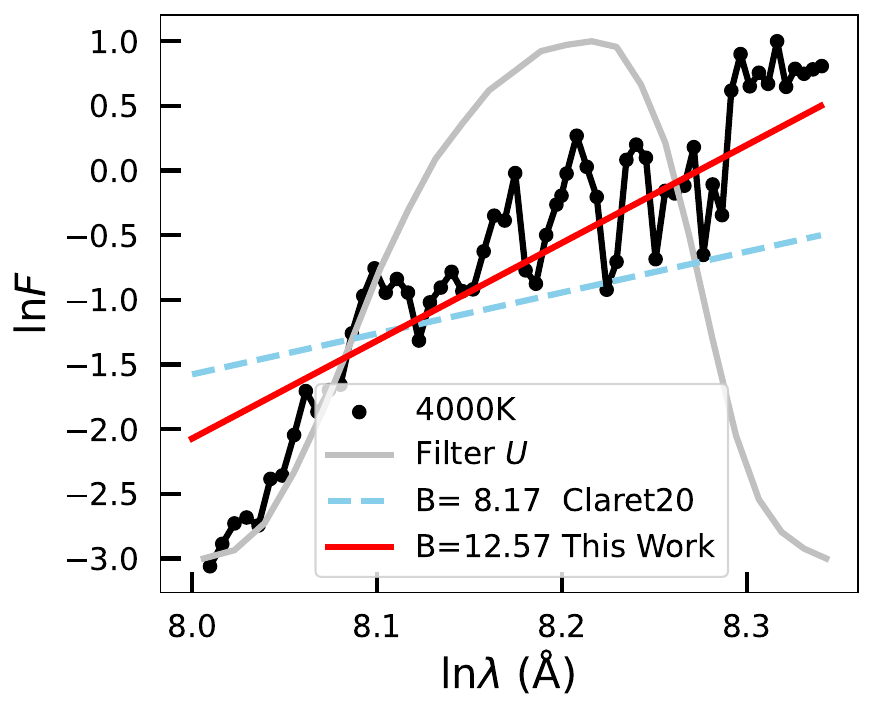}
    \includegraphics[scale=0.4]{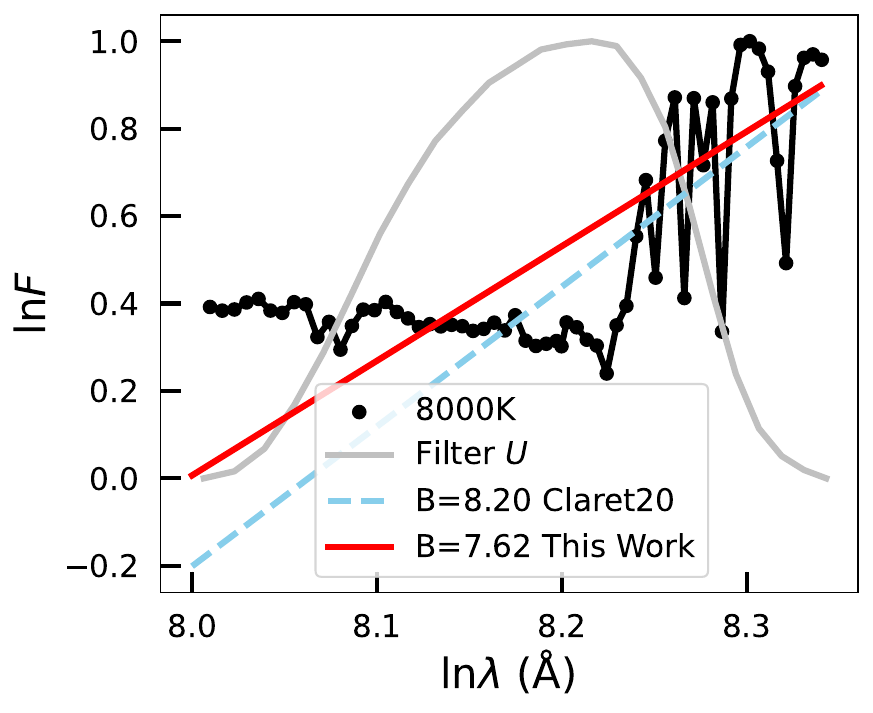}
    \includegraphics[scale=0.4]{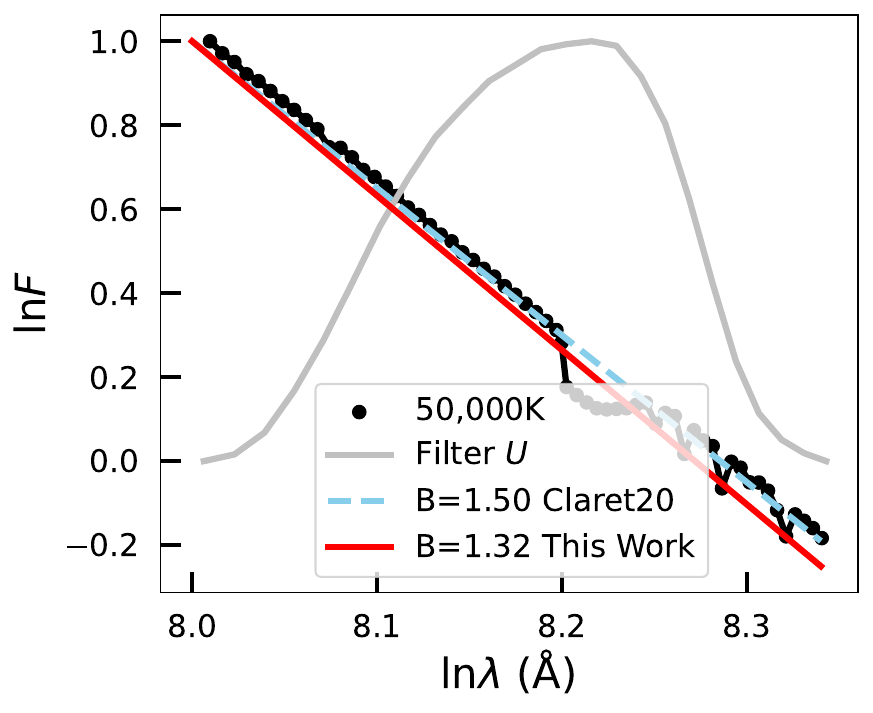}
    \includegraphics[scale=0.4]{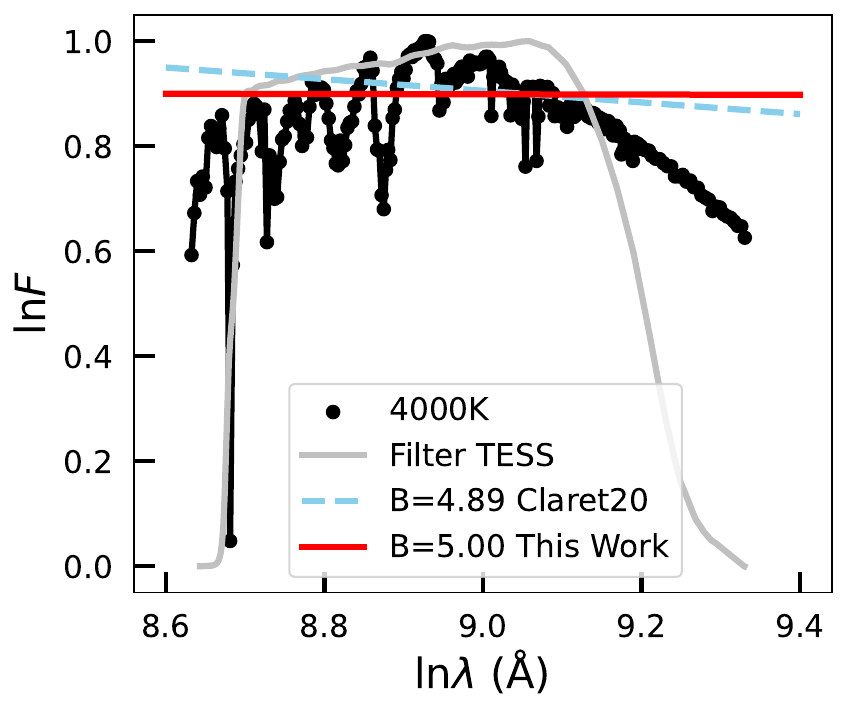}
    \includegraphics[scale=0.4]{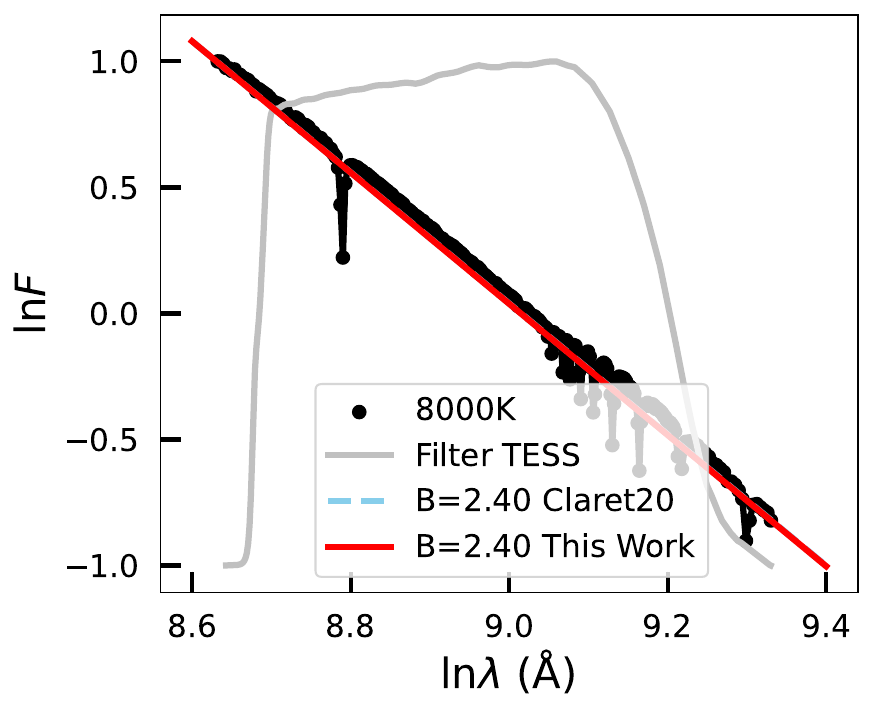}
    \includegraphics[scale=0.4]{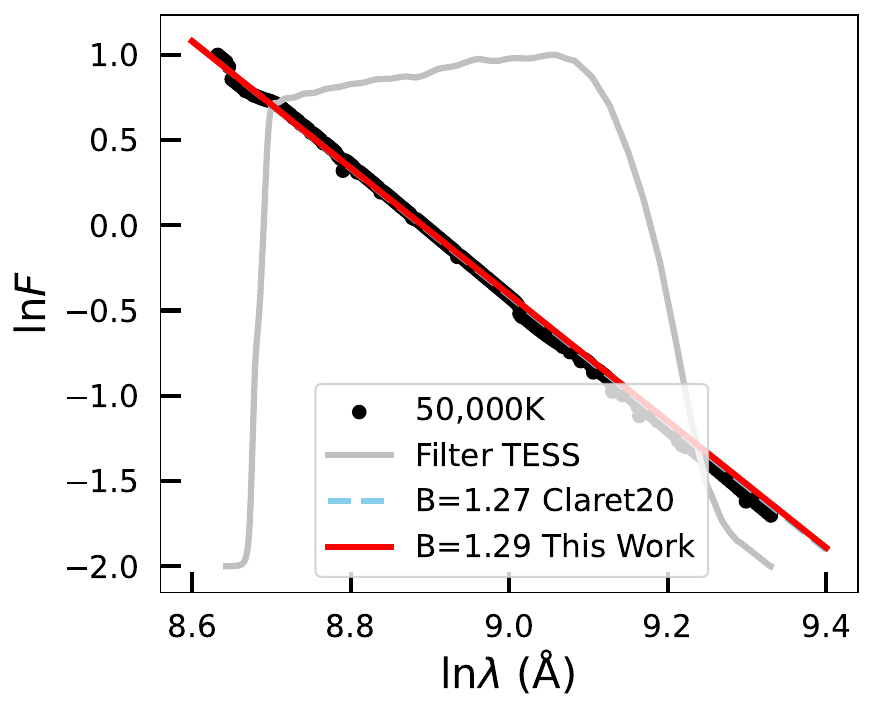}
    
    \caption{Predicted power-law profile inferred from the beaming factors in this work (red-solid line) and \cite{claret} (cyan-dashed line). We note that both profiles are shifted in arbitrary amounts to match the synthetic spectra from \texttt{ATLAS9} with fixed log\,$g = 5.0$, Solar metallicity, and effective temperature labeled in the bottom-left corner. The upper three panels show results for the $U$-band, and the bottom three panels are for the TESS band. Transmission curves for both bands are overplotted with gray-solid lines.}
\label{fig:compare_index}
\end{figure*}

\vskip 1.5cm
\section{The Doppler Beaming Factor}
According to relativistic theory, the ratio between the flux $F(\nu)$ and the cube of the frequency $\nu^{3}$ should remain constant across different reference frames (see Equation 4.111 in 
\citealt{theory}). 

By adopting a relativistic Doppler shift $\nu=D\,\nu_{0}$, we obtain:
\begin{equation}
 F(\nu)_{\rm{obs}}=F(\nu/D)_{0}D^3\text{\ ,}
    \label{eq:niu}
\end{equation}
Here, $F(\nu)_{\rm{obs}}$  refers to the observed flux and the subscript 0 denotes the corresponding quantities in the absence of the relative motion. $D$ is given by:
\begin{equation}
    D= \frac{1}{\gamma(1+\frac{v_{r}}{c})} , 
    \label{eq:D}
\end{equation}
and:
\begin{equation}
    \gamma= \frac{1}{\sqrt{1-\frac{{v}^2}{c^2}}}\text{\ .}
        \label{eq:gamma}
\end{equation}
Here, ${v_{r}}$ refers to the stellar radial velocity and $v$ refers to stellar space velocity. 
The monochromatic beaming factor $B$, as a function of $v_{r}$, can be derived at a given frequency $\nu_{i}$:
\begin{equation}
    B(v_{r})= \frac{\partial F(\nu_{i})_{\rm{obs}}}{\partial \frac{v_{r}}{c}}\text{.}
    \label{eq:common}
\end{equation}

\subsection{The Linear Approximation: A Constant Beaming Factor}
\label{sec:powerlaw}
To search for a binary or a planet plus host-star systems using beaming effects \citep{BEAM_planet}, a further refinement of  Equation \ref{eq:niu} is required, based on two assumptions: 1) the low-speed approximation, where the stellar space velocities $v$ are far from the speed of light; and 2) the stellar spectra are represented by power-law profiles: $F \propto \nu^\alpha$.
Under these two assumptions, $D$ is then equal to $1-\frac{v_r}{c}$; thus, Equation \ref{eq:niu} can be re-written as:
\begin{equation}
    F(\nu)_{obs} = F(\nu)_0\left(1-(3-\alpha)\frac{v_r}{c}\right) \text{.}
    \label{eq:alpha1}
\end{equation}
Here, the beaming flux exhibits a linear dependency on the radial velocity $v_r$; thus, the beaming factor becomes a constant value of $(\alpha-3)$, as defined in Equation\,\ref{eq:common}.
We note that this factor differs from that of \cite{BEAM_planet} by a minus sign, simply because they define redshift in the opposite direction from the usual convention.

Naturally, under the power-law approximation, Equation \ref{eq:alpha1} can also be written in wavelength space as: 
\begin{equation}
    F(\lambda)_{\rm{obs}} = F(\lambda)_0\left(1-B(\lambda)\frac{v_{r}}{c}\right)\text{,}
    \label{eq:beaming2}
\end{equation}
and:
\begin{equation}
    B(\lambda)=5+ \frac{d \rm{ln}\emph{F}(\lambda)_0} {d \rm{ln}\lambda}\text{.}
    \label{eq:B}
\end{equation}

\subsection{Applications to Widely Used Filter Systems}

To compute beaming factors for the most widely used photometric
systems, \cite{BEam_obs_1} and \cite{claret} modified the monochromatic factor $B(\lambda)$ to the photon-weighted bandpass-integrated beaming factor $\bar{B}$, incorporating filter transmission curves $T(\lambda)$ and synthetic spectra $f(\lambda)$:
\begin{equation}
    \bar{B} = \frac{\int_{\lambda_1}^{\lambda_2}{T(\lambda)\lambda B(\lambda)f(\lambda)d\lambda}}{\int_{\lambda_1}^{\lambda_2}{T(\lambda)\lambda f(\lambda)d\lambda}}\text{,}
    \label{eq:claret}
\end{equation}
where $\lambda_1$ and $\lambda_2$ denote the lower and upper wavelengths for a specified filter.

Following Equation \ref{eq:claret}, \cite{claret} presents the first publicly available beaming-factor table for hot white dwarf stars (DA, DB, and DBA), main-sequence stars, and giant stars in commonly adopted broad-band systems, including the SDSS, $UBVRI$, HiPERCAM, $Kepler$, TESS, and $Gaia$ bands. 
The white dwarf atmospheric models from \cite{WD_model_claret} and models from \cite{claret} for
main-sequence and giant stars are adopted in their calculations. 
We note that their calculations didn't consider the reddening effect, which cannot be ignored for disk and bulge stars. 
\begin{table*}
    \centering
    \caption{\textbf{Synthetic Spectral Libraries Adopted in this Work} }
    \centering
    \begin{threeparttable}
    \begin{tabular*}{\textwidth}{@{\extracolsep{\fill}}cccccc}
    \hline
    Library&Reference &Resolving power &$T_{\rm{eff}}$ (K)& log $g$   &Metallicity \\

    \hline
    {\tt PHOENIX}$^a$&\cite{PHOENIX}& 500,000 & [3000, 12,000] & [0.0, 6.0]  & [$-4.0$, +1.0]$^e$\\
    {\tt ATLAS9}$^b$ &\cite{atlas9} & 150-300& [3000, 50,000]  & [0.0, 5.0]  & [$-4.0$, +0.5]$^e$ \\
    {\tt TMAP}$^c$ & \cite{tmap_lib} & $\geq$10,000& [50,000, 190,000] & [5.0, 9.0] & [0.0, +1.0]$^f$ \\
    {\tt TLUSTY}$^d$ & \cite{TLUSTY_O,TLUSTY_B} & 1800$^g$ & [15,000, 55,000] & [1.75, 4.75] & [0.0, 2.0]$^h$\\
    \hline
    \end{tabular*}      
    
    \label{tab:tabel}
     \begin{tablenotes}
        \footnotesize
        \item[$a$] The average steps in $T_{\rm{eff}}$, log $g$, and metallicity are 100 K, 0.5 dex, and 0.5 dex, respectively.
        \item[$b$] The average steps in $T_{\rm{eff}}$, log $g$, and metallicity are 250 K, 0.5 dex, and 0.5 dex, respectively.
        \item[$c$] The average steps in $T_{\rm{eff}}$, log $g$, and metallicity are 10,000 K, 0.5 dex, and 0.1 dex, respectively.
        \item[$d$] The average steps in $T_{\rm{eff}}$ and log $g$ are 1000 K and 0.25 dex, respectively. The metallicity grid is represented by 0.0, 0.1, 0.2, 0.5, 1.0, and 2.0, respectively.
        \item[$e$] Here, metallicity is taken to be [Fe/H].
        \item[$f$] Here, metallicity refers to the mass ratios between H and He.
        \item[$g$] The resolving power drops to 450 for metal-free cases.
        \item[$h$] Here, metallicity refers to Z/Z$_\sun$.
    \end{tablenotes}
  \end{threeparttable}
\end{table*}

\vskip 1cm
\subsection{A New Method: Back to the Definition}
\label{sec:method}

The constant beaming factor defined in previous studies is based on the power-law approximation, which actually deviates significantly from real stellar spectra, especially in the blue range of optical spectra. This leads us to consider whether or not a linear relationship between observed flux and radial velocity is suitable for all types of stars observed in various filter systems, which has never been tested.
We thus propose a new method to calculate the beaming flux at various radial velocities and then compute beaming factors based on the direct definition given in Equation \ref{eq:common}.

In comparison to the previous approach, the new technique has two advantages: 1) it facilitates the examination of whether a constant beaming factor adequately describes the relation between the observed flux and radial velocity; and 2) the calculation is straightforward and applicable to any kind of spectra; for example, those regions heavily affected by metallic absorption or even the entire optical bands covered by molecular bands for cool stars. In the following subsections, we provide a detailed introduction to the new method for computing the beaming factor, as well as a method to check its validity as a constant.

\vskip 1cm
\subsubsection{Synthetic Beaming Flux and Beaming Factor}
\label{sec:flux&bf}

First, using theoretical stellar spectra, the monochromatic beaming flux at a given radial velocity can be determined by Equation \ref{eq:niu}:
\begin{equation}
    F(\lambda)_{\rm{beam}}=F(D\lambda)_0 D^5\text{.}
    \label{eq:lambda}
\end{equation}

For filters with transmission curve $T(\lambda)$ and targets with color excess $E (B-V)$, the synthetic beaming flux $\bar{F}_{\rm{obs}}$ observed by energy-counting detectors can be computed with:

\begin{equation}
    \bar{F}_{\rm{obs}}  = \frac{\int_{\lambda_1}^{\lambda_2}F(\lambda)_{\rm{beam}}R(\lambda)T(\lambda)d\lambda}{\int_{\lambda_1}^{\lambda_2}{T(\lambda)d\lambda}}\text{,}
    \label{eq:filter}
\end{equation}
where $R(\lambda)$ is the reddening term:
\begin{equation}
    R(\lambda)=10^{-\frac{k_{\lambda}E(B-V)}{2.5}} \text{,}
\end{equation}
where $k_{\lambda}$ represents the reddening coefficient predicted from the $R_V = \frac{A_{V}}{E (B-V)} = 3.1$, the Fitzpatrick extinction law \citep{F19}.

For a specific beaming binary with a circular orbit $e = 0$, a semi-amplitude $K$ of a radial velocity curve and a reddening value of $E (B-V)$, the beaming factor of a given filter with transmission curve $T (\lambda)$ can be directly calculated by Equation~\ref{eq:common}: 
\begin{equation}
B_{X} \left(K, E (B-V)\right) = \frac{\partial \bar{F}_{\rm obs}}{\partial \frac{v_{r}}{c}} ( - K \le v_r - v_{\rm{sys}} \le K)\text{,}
\label{eq:BF}
\end{equation}
where $X$ indicates the name of a given filter and $\bar{F}_{\rm obs}$ can be computed from Equation~\ref{eq:filter}.

The method can also work for eccentric orbits ($e \neq 0$), by revising the upper and lower boundaries of radial velocity using the binary orbital parameters \citep{RV_derive}.
A dedicated Python code, {\tt BeamingFactor}, has been developed to calculate the beaming factor based on Equation~\ref{eq:BF}. 
By default, the code calculates $\bar{F}_{\rm obs}$ from $-K$ to $K$ in steps of 1\%$K$, or $2.5$\,km\,s$^{-1}$ when $K \ge 250$\,km\,s$^{-1}$.
The beaming factor is then determined from the slope of a linear fit between $\bar{F}_{\rm obs}$ and $v_r$.
The code accommodates both the well-known filter systems integrated within the code and any passbands provided by the user.
For synthetic spectra, this code can directly read spectra files form several widely used libraries: {\tt PHOENIX} \citep{PHOENIX}, {\tt ATLAS9} \citep{atlas9}, and {\tt TAMP} \citep{tmap_lib}, generated with the 
codes {\tt PHOENIX} \citep{PHOENIX_ORI}, {\tt ATLAS9} \citep {kurucz1,kurucz2}, and 
{\tt TAMP} \citep{tmap0,tmap1,tmap2}, respectively.
Additionally, users can upload their own spectral libraries.
The code is made publicly available at \href{https://doi.org/10.5281/zenodo.13049419}{Zenodo} \citep{beamingfacotr}, and a version under continuous development is available at \url{https://github.com/shift-method/beamingfactor}.

\subsubsection{Beyond the Linear Approximation}
\label{sec:alp}

In contrast to \cite{claret}, our procedure offers the capability to assess whether a constant beaming factor sufficiently captures the relationship between the observed flux and radial velocity.
In other words, our method can work effectively even if the relationship is not linear, as assumed in previous studies.
To accomplish this, we define the $D_{5}$ index, which counts the fraction of data points deviating from the linear fit by $\frac{\delta K}{K} \Delta F$, with $\delta K = 5\%K$.
Here $\Delta F$  is the maximum flux change within an orbit: $\bar{F}_{\rm obs} (v_r = v_{\rm sys} - K) - \bar{F}_{\rm obs} (v_r = v_{\rm sys}+K)$.
Similarly, we can define $D_{1}$ and $D_{2}$ indexes for different types of studies.
If the deviation from linearity is significant, let's say $D_5 = 10$\% (users can define this fraction according to their own requirements), the constant beaming factor will fail to accurately convert the observed flux variation to radial-velocity variation.

Figure~\ref{fig:applicable} presents two examples, both asssuming a binary system in a circular orbit with semi-amplitude $K = 300$\,km\,s$^{-1}$. 
In the left panel, the primary of the binary is a Solar-type extremely metal-poor star ($T_{\rm eff} = 5600$\,K, log\,$g = 4.0$, and [Fe/H]\,=\,$-3.0$) with zero reddening, $E (B-V) = 0$.
The beaming flux at the $B$-band exhibits a perfectly linear correlation with radial velocity, and the $D_5$ index is zero.
However, in the right panel, clear deviations from a linear fit are evident between the beaming flux in the $B$-band and radial velocity for a metal-rich cool giant ($T_{\rm eff} = 2600$\,K, log\,$g = 2.0$, and [Fe/H]\,=\,$+0.5$) with moderate reddening, $E (B-V) = 0.3$. This deviation is a natural result of the strong absorption within the $B$ band, which causes the spectra to deviate significantly from a power-law profile.
The $D_{5}$ index is as high as 21\% and $D_{2}$ reaches as high as 69\%.
The radial velocity-dependent beaming factors are required to convert the observed flux variations into a radial-velocity curve.  This can be done by our code, if needed, while previous methods cannot achieve this.

As stated earlier, our method is more straightforward for computing the beaming factor than the former method based on the power-law assumption for stellar spectra. This is particularly true for the blue bands and cool stars, whose spectra are dominated by absorption features. Figure~\ref{fig:compare_index} shows the comparison between the real synthetic spectra and spectral indexes inferred from the beaming factors computed from our method and from \cite{claret}, which assume a power-law profile for stellar spectra.
As expected, the spectral slopes predicted by our beaming factor are in excellent agreement with the synthetic ones across all effective-temperature ranges (i.e., $T_{\rm eff}$ from 4000\,K to 50,000\,K) in both the blue and optical bands. On the contrary, the slopes given by the beaming factors from \cite{claret} exhibit significant deviations in the 
$U$-band spectra across all effective-temperature ranges and moderate deviations even for cooler temperatures (i.e., $T_{\rm eff} = 4000$\,K) in the optical band.
These comparisons clearly demonstrate the superiority of our method for calculating the beaming factor.

\section{Beaming Factor Tables}

Similar to \cite{claret}, a comprehensive table of beaming factors is calculated using the {\tt BeamingFactor} code developed in this study.
The factors are computed for binaries with circular orbits and a semi-amplitude $K = 300$~km~s$^{-1}$, which are applicable to most known beaming binaries \citep{KPD1946,BEER2,HD265435,beaming10,beaming11}. The table also includes $D_1$, $D_2$ and $D_5$ to help evaluate linearity and precision.  The results are partly summarized graphically in Figures 3 and 4.
We note that the results also hold very well for binaries with smaller or larger values of $K$.
However, if one is interested in very blue passbands (e.g., $U$) or cool stars, it is  better to calculate them individually using the {\tt BeamingFactor} code with the exact values of the binary orbital parameters.

\begin{figure*}
    \centering
    \includegraphics[scale=0.39]{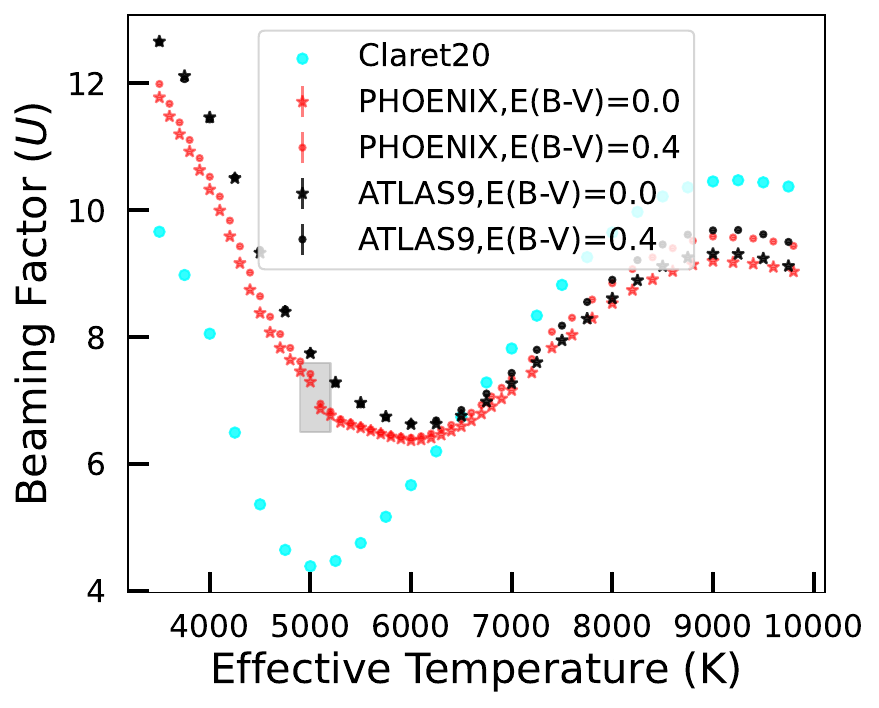}
    \includegraphics[scale=0.39]{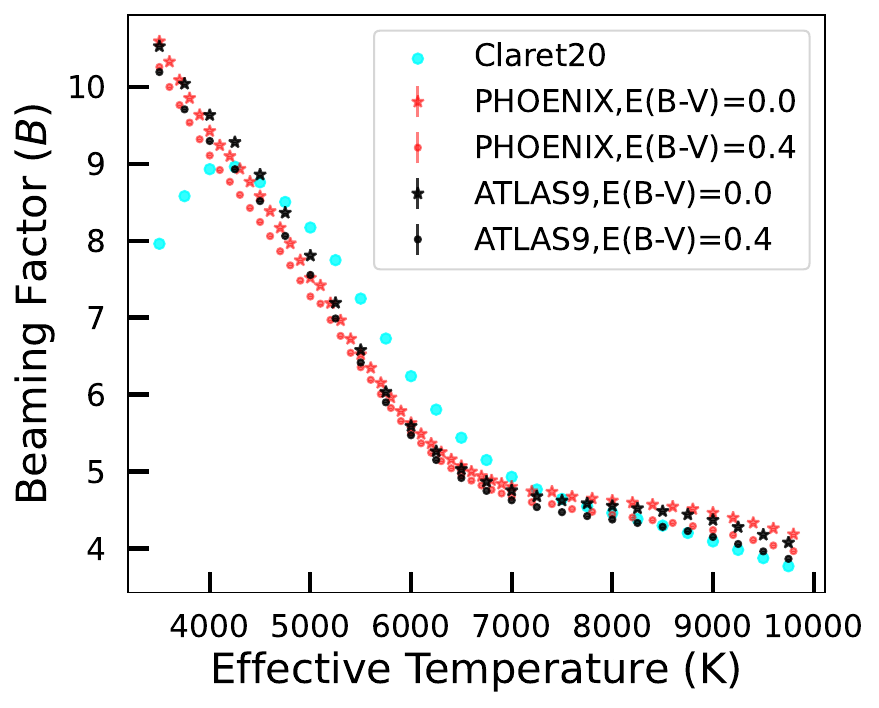}
    \includegraphics[scale=0.39]{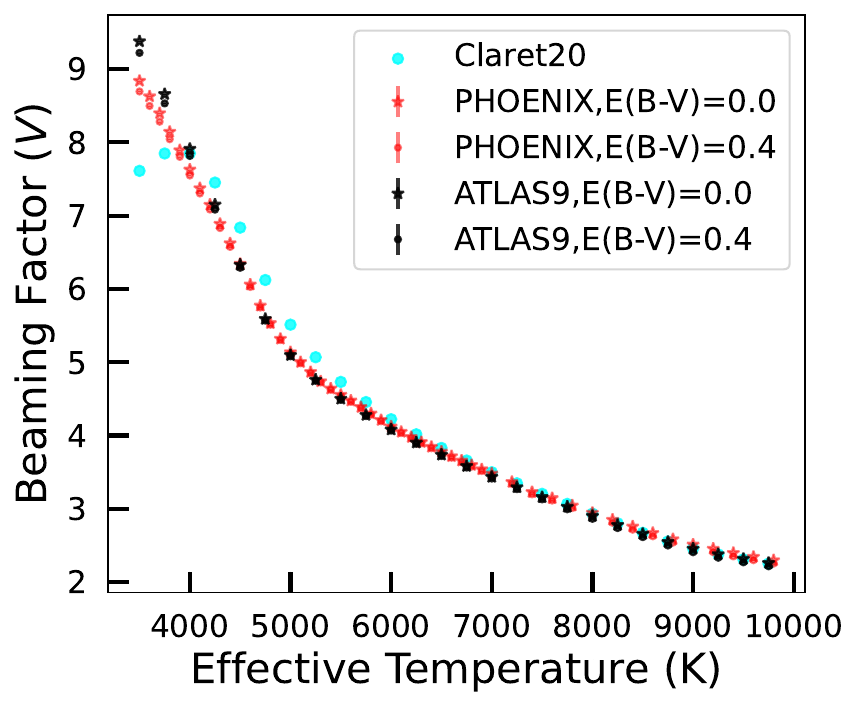}
    \includegraphics[scale=0.39]{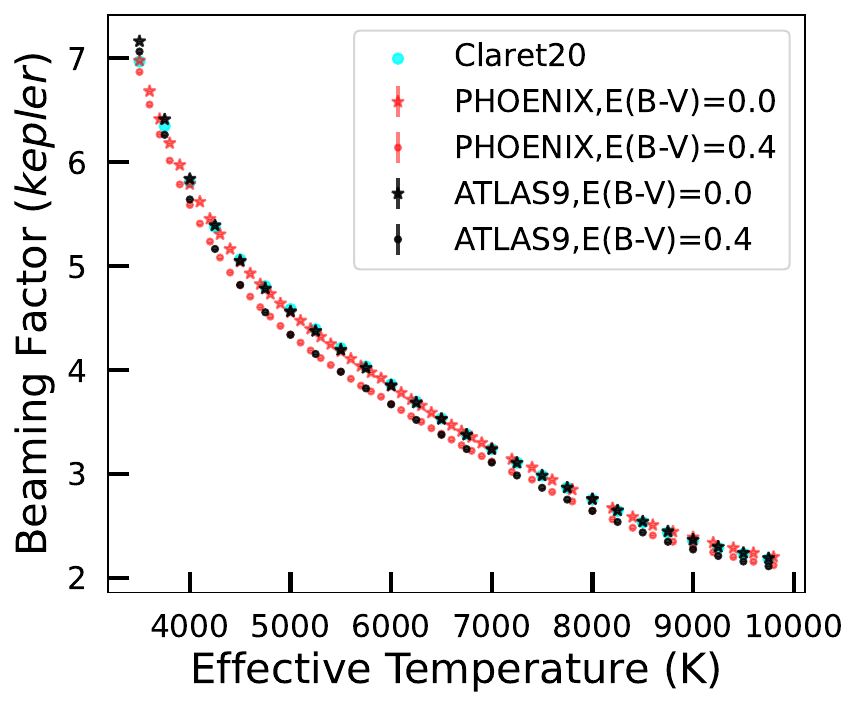}
    \includegraphics[scale=0.39]{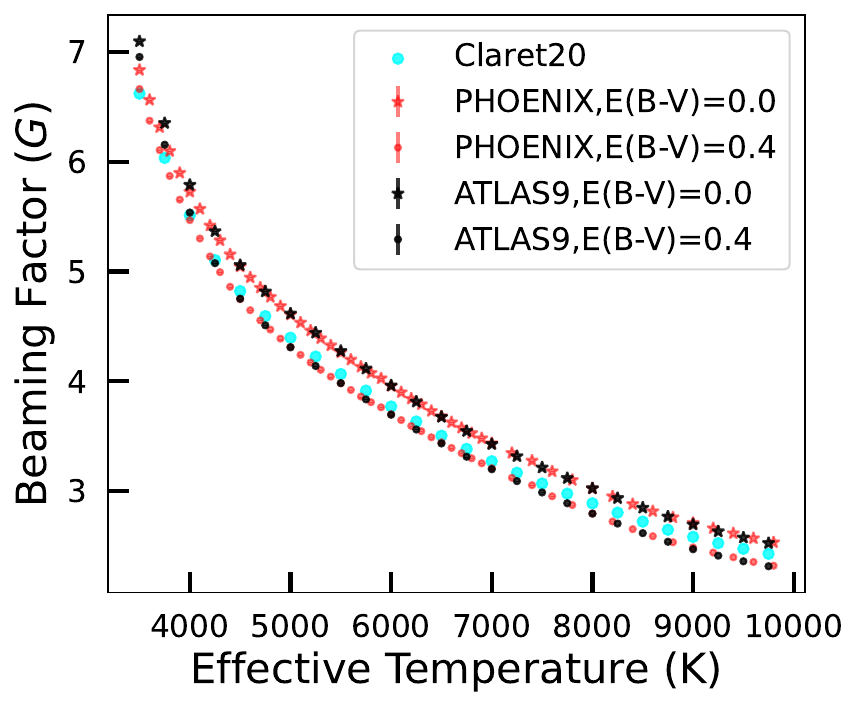}
    \includegraphics[scale=0.39]{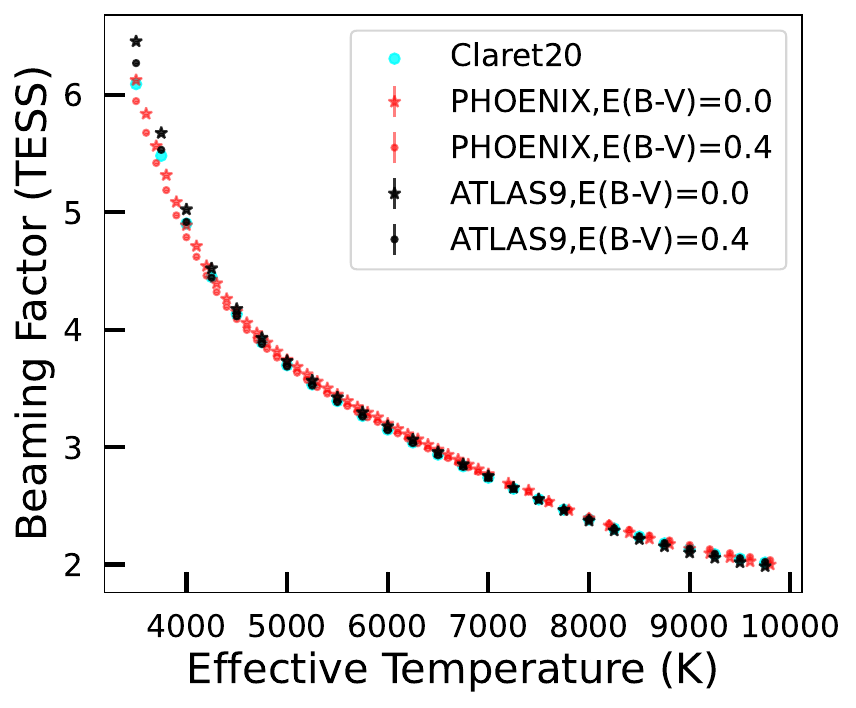}

    \caption{Beaming factors for the $U$, $B$, $V$, $Kepler$, $Gaia$ $G$, and \texttt{TESS} filters, as a function of effective temperature ranging from 3500\,K to 10,000\,K. All results are computed with ATLAS9 \citep{atlas9} and PHOENIX \citep{PHOENIX} synthetic spectra with Solar metallicity, log\,$g$ = 4.5. The results from \cite{claret} are also included for comparison. 
    Readers who wish to use beaming factors within the gray areas should carefully examine the corresponding spectra to ensure no artificial features are present. The decreasing trend observed in all filters except for {\it U} reflects the evolution of the spectral slope with $T_{\rm eff}$. The `U' shape in the {\it U} filter indicates an additional influence from the Balmer jump within the {\it U} filter, which otherwise would increase the slope.}
\label{fig:compare}
\end{figure*}

\begin{figure*}
    \centering
    \includegraphics[scale=0.39]{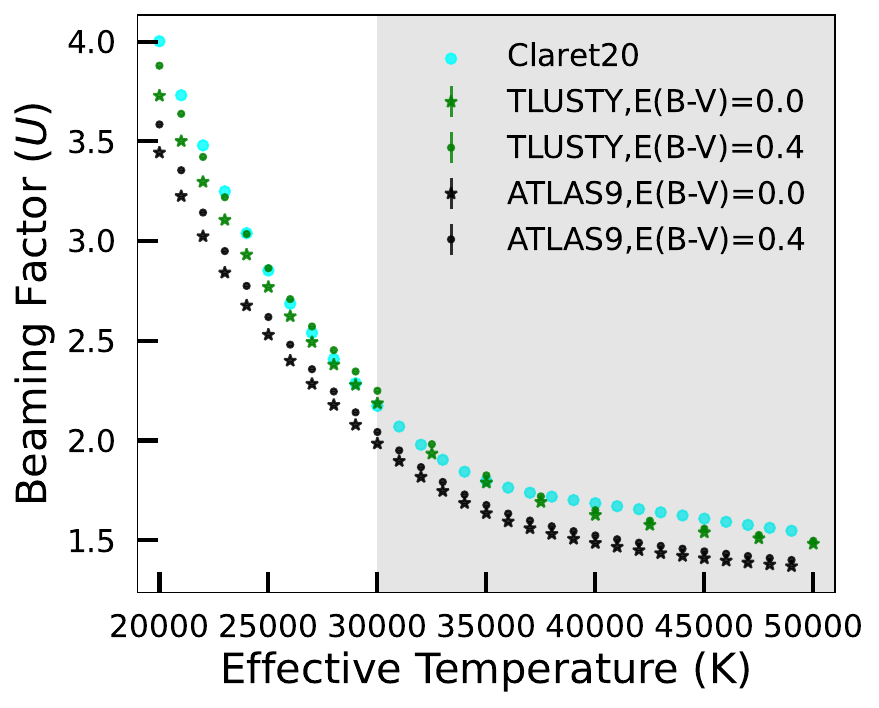}
    \includegraphics[scale=0.39]{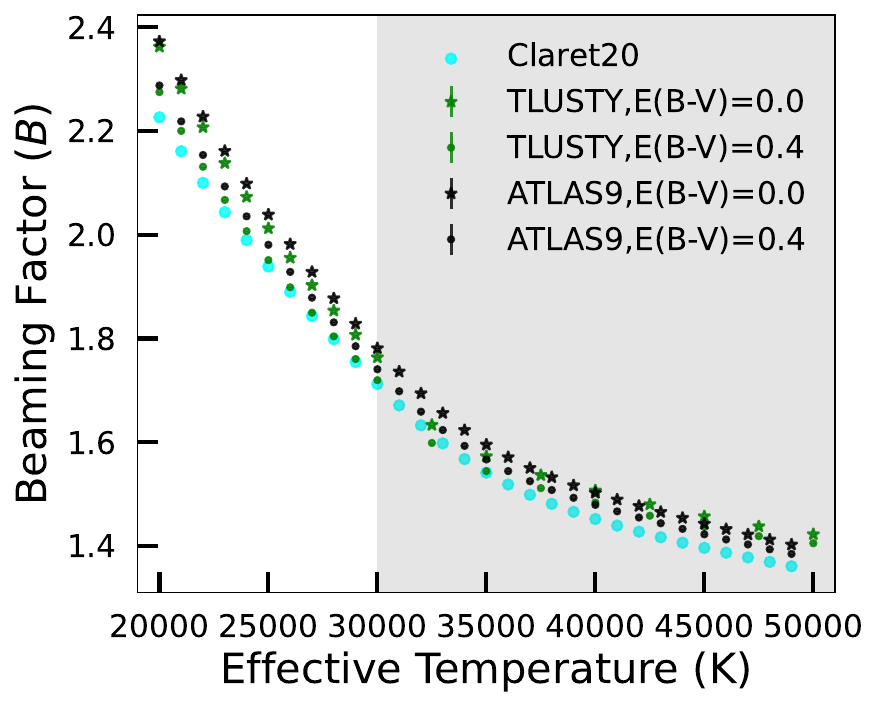}
    \includegraphics[scale=0.39]{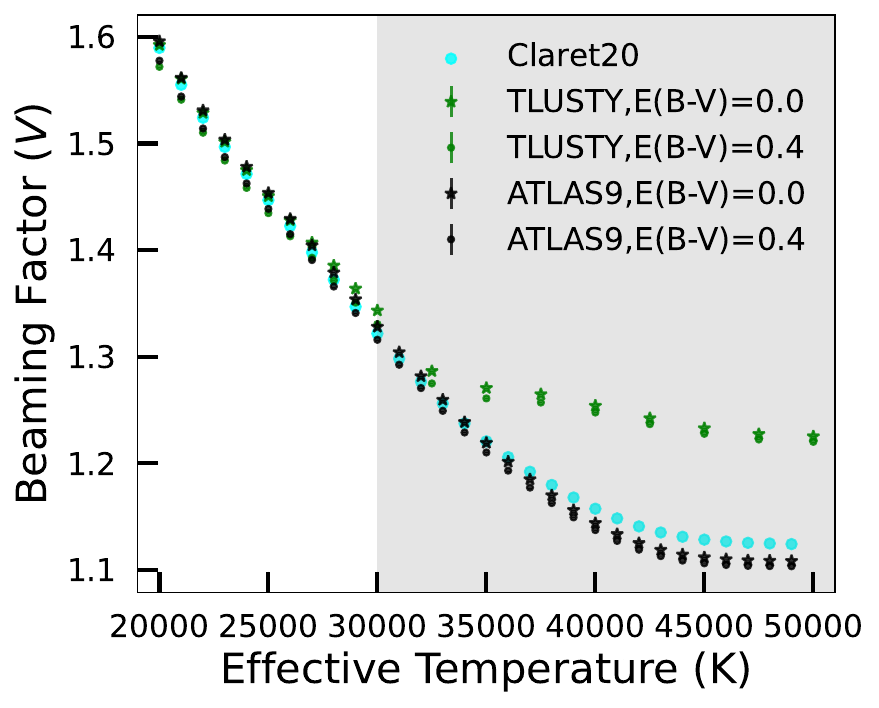}
    \includegraphics[scale=0.38]{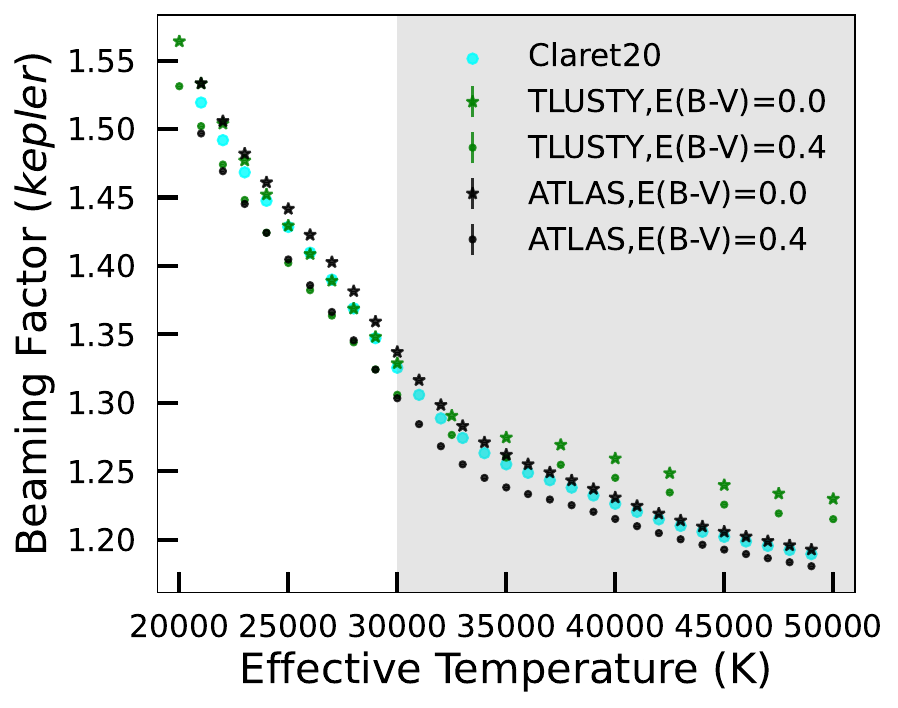}
    \includegraphics[scale=0.38]{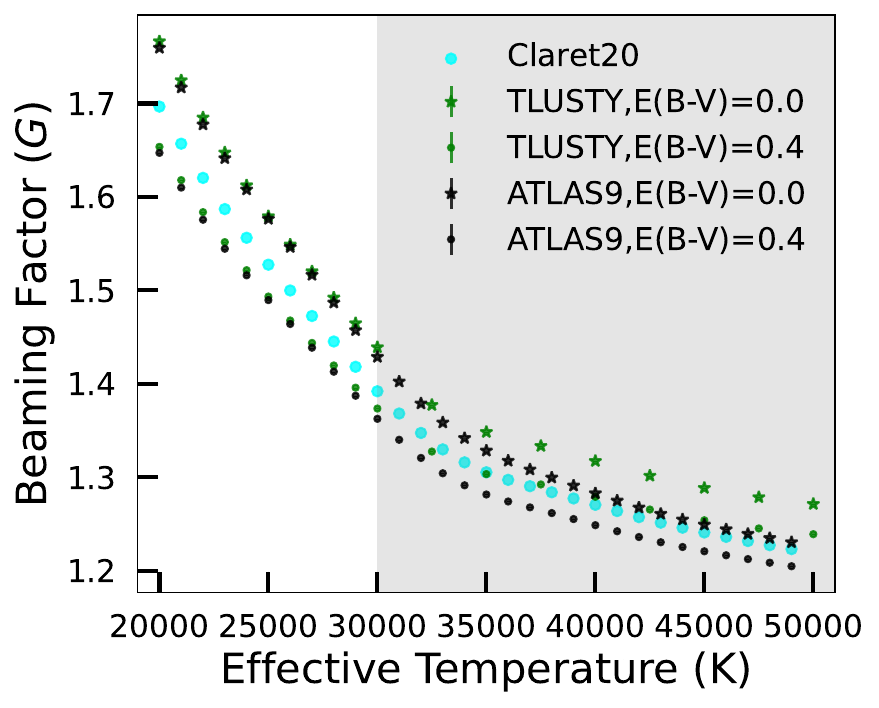}
    \includegraphics[scale=0.38]{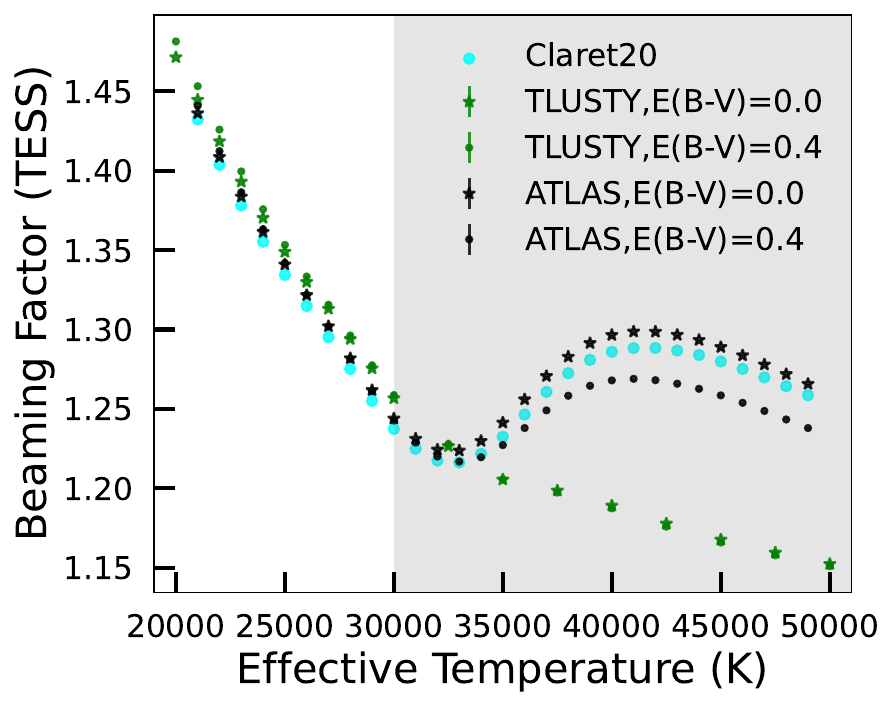}    
    \caption{Similar to Figure~\ref{fig:compare}, but for effective temperature ranging from 
    20,000\,K to 50,000\,K and with synthetic spectra libraries {\tt TLUSTY} \citep{TLUSTY_O,TLUSTY_B} and {\tt ATLAS9} \citep{atlas9}. The factors computed with {\tt ATLAS9} present a ``U" shape in the TESS filter, which originates from artificial spectral features near 5881\,\AA~ (see Figure \ref{fig:5800A} in appendix for details). We thus recommend to visually check the spectra when computing beaming factors for hotter stars with $T_{\rm eff} \ge$~30,000~K (see gray areas).}
\label{fig:compare_h}
\end{figure*}

\subsection{Parameter Coverage}

The spectral libraries we adopted are {\tt PHOENIX} \citep{PHOENIX}, {\tt ATLAS9} \citep{atlas9}, {\tt TLUSTY OSTAR2002 and BSTAR2006} \citep[hereafter {\tt TLUSTY};][]{TLUSTY_O,TLUSTY_B}, and {\tt TAMP} \citep{tmap_lib}. 
Table\,1 summarizes the resolving power and atmospheric parameter coverages of the four libraries. Our computations incorporate all available spectra files, so the factor tables follow the same parameter grids as these synthetic spectra libraries.
In short, both {\tt PHOENIX} and {\tt ATLAS9} work very well for normal main-sequence, subgiant, and giant stars over wide metallicity ranges, while {\tt ATLAS9} provides spectra for hot stars with effective temperature as high as 50,000\,K.
{\tt TLUSTY} is dedicated for OB stars with $T_{\rm{eff}}$ ranging from 15,000 K to 55,000 K.
For white dwarf stars and hot subdwarf stars, the {\tt TMAP} libraries are recommended, as they computed an extended atmosphere grid of H+He composition models with different H/He abundance ratios.
For more details about these four libraries, we refer the interested reader to \cite{PHOENIX}, \cite{atlas9}, \cite{TLUSTY_O,TLUSTY_B}, and \cite{tmap_lib}.

Different choices of reddening values\footnote{The grid contains values of $E (B-V)$: 0.0, 0.05, 0.10, 0.15, 0.20, 0.25, 0.30, 0.40, 0.60, 0.80, 1.00, 1.25, and 1.5.} are also considered for calculating the beaming factors.
In total, we calculated beaming factors for six filter systems, including SDSS ($ugriz$), Johnson-Cousins ($UBVRI$), $Gaia$ ($G,G_{\rm BP},G_{\rm RP}$), $Kepler$, TESS, and CSST ($NUV$ and $ugrizy$).

\subsection{Results and Comparison}
\label{sec:discuss}
The results for beaming factors are made publicly available at \href{https://doi.org/10.5281/zenodo.13049419}{Zenodo} \citep{beamingfacotr}.
For the listed filter systems, the beaming factors vary between 1 and 15. Consequently, the Doppler-beaming signal is typically on the order of a few thousandths or less when $K$ is around 100 km~s$^{-1}$. This tiny signal demands highly precise photometry to detect, as it is much smaller compared to other effects such as ellipsoidal modulation and reflection, which can reach several tens of percent. Although its amplitude is small, the Doppler-beaming signal is distinguishable from these other effects, enabling clear separation. To facilitate this, a dedicated program named {\tt BEER} was developed by \cite{BEER0} for the photometric detection and analysis of these signals.
However, we note that another hard-to-distinguish factor is the presence of varying stellar spots, which are commonly found in M dwarf stars with strong magnetic activity.

In Figures~\ref{fig:compare} and \ref{fig:compare_h}, we present a comparison between the beaming factors computed in this study and those calculated by \cite{claret}.
Overall, for factors computed with {\tt ATLAS9}, as also adopted  by \cite{claret}, the results exhibit consistency, with one notable exception: for cool stars with $T_{\rm eff} \le 7000$\,K in the $U$-band, the values reported by \cite{claret} are 20\% to 40\% lower than those obtained in our study.
Such differences between factors will result in different values of the photometric semi-amplitude radial velocity, $K$.
Even for hot stars with $T_{\rm eff} \ge 20,000$\,K, the deviations remain at 10\% in the $U$-band.
This discrepancy is easily understood, since their method has problems capturing the real slope of the spectra in the blue region of stars, especially for those that are cooler than $8000$~K (see Figure~\ref{fig:compare_index}).
Moreover, we note that reddening can cause variations of a few to ten percent in the derived beaming factors, especially for blue bands.
This effect has now been properly accounted for in our tables, with different choices of $E(B-V)$ ranging from 0.0 to 1.5 provided.

In contrast to the substantial discrepancy between the {\tt ATLAS9} results from this work and those of \cite{claret}, the beaming factors computed with different models exhibit only minor differences, typically within 1\%.
However, significant discrepancies (as large as 20\%) are detected for $T_{\rm{eff}}$ higher than 30,000\,K in $V$ and TESS filters (see right panels of Figure \ref{fig:compare_h}). 
Such deviations may arise from two factors.  First, the models use different assumptions.
For example, {\tt ATLAS9} generates synthetic spectra assuming local thermodynamic equilibrium (LTE), whereas {\tt TLUSTY} incorporates non-LTE (NLTE) effects.
We thus recommend the beaming factors calculated by {\tt TLUSTY}, given that NLTE effects are important when modeling the atmosphere of hotter stars.
Second, modeling synthetic spectra of hotter stars may encounter numerical convergence issues, leading to unexpected artificial features in the resulting spectra.
For example, the beaming factors for the TESS filter display a ``U" shape as a function of $T_{\rm eff}$ when using {\tt ATLAS9} spectra, whereas the results from {\tt TLUSTY} spectra show a decreasing trend, as expected.
This unexpected ``U" feature is actually caused by artifacts around 5881~\AA\, in the {\tt ATLAS9} spectra for $T_{\rm eff}$ above 30,000 K (see Figure \ref{fig:5800A} in the appendix).
We therefore recommend that readers visually inspect the spectra when using beaming factors for hotter stars with $T_{\rm eff} \ge 30,000$~K.

\section{Summary}

In this study, we present a novel approach for computing the beaming factor for binary systems based on its direct definition. This method is superior to previous approaches that relied on the assumption of a power-law profile for stellar spectra. Our method offers several advantages, particularly in two aspects: 1) calculating the factors straightforwardly, especially for blue bands and cool stars whose spectra are dominated by absorption features; and 2) evaluating whether a constant factor can accurately describe the relationship between beaming flux and radial velocity, which previous methods assumed.

We have developed the Python code {\tt BeamingFactor} to derive beaming factors for binary systems with known orbital parameters with any photometric bands. Using this code, we provide a table of beaming factors for users to quickly reference. This table is computed for widely used photometric filter systems, including SDSS ($ugriz$), Johnson-Cousins ($UBVRI$), $Gaia$ ($G, G_{\rm BP}, G_{\rm RP}$), $Kepler$, TESS, and CSST ($NUV$ and $ugrizy$), covering main-sequence, subgiant, and giant stars, as well as hot subdwarf and white dwarf stars.
Both the code and table are archived and public available at \href{https://doi.org/10.5281/zenodo.13049419}{Zenodo} \citep{beamingfacotr}. For the code under continuous development, please refer to \url{https://github.com/shift-method/beamingfactor}.

\begin{acknowledgements}
\section*{acknowledgements}
We would like to thank the referee for helpful comments.
This work acknowledges support from the National Natural Science Foundation of China (NSFC) for grants No. 11933004 and No. 11988101.
YH acknowledges support from the National Key Basic R\&D Program of China via 2023YFA1608303 and the China Manned Space Project with No. CMS-CSST-2021-A08.
JFL acknowledges support from the
New Cornerstone Science Foundation through the New Corner-
stone Investigator Program and the XPLORER PRIZE.  T.C.B. acknowledges partial support for this work from grant PHY 14-30152; Physics Frontier Center/JINA Center for the Evolution of the Elements (JINA-CEE), and OISE-1927130: The International Research Network for Nuclear Astrophysics (IReNA), awarded by the US National Science Foundation. 
\end{acknowledgements}

\bibliography{refer.bib}{}
\bibliographystyle{aasjournal}
\appendix
\setcounter{table}{0}   
\setcounter{figure}{0}
\renewcommand{\thetable}{A\arabic{table}}
\renewcommand{\thefigure}{A\arabic{figure}}
\section{Synthetic spectra comparisons between {\tt ATLAS9} and {\tt TLUSTY}}
\begin{figure}[h]
    \centering
    \includegraphics[width=0.95\linewidth]{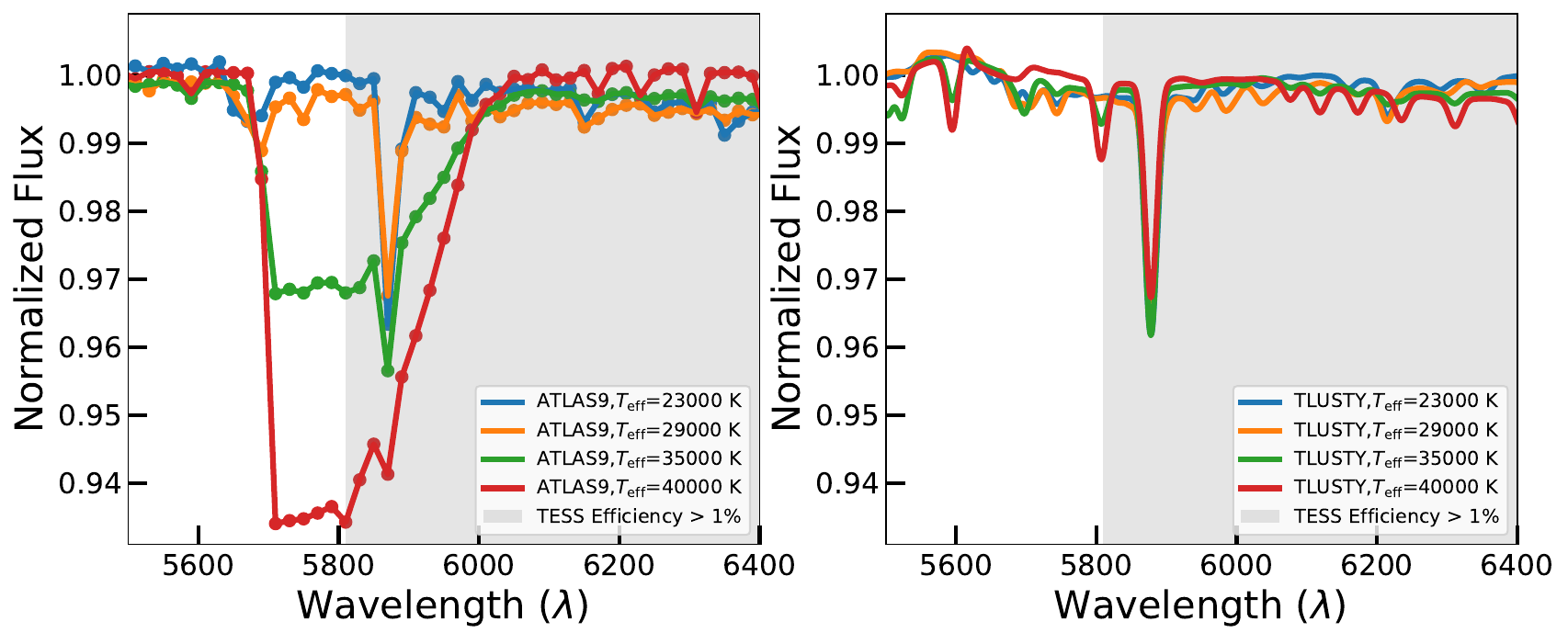}
    \caption{Normalized spectra from {\tt ATLAS9} (left panel) and {\tt TLUSTY} (right panel) are shown for $T_{\rm eff}$ of 23,000, 29,000, 35,000, and 40,000~K. All spectra have Solar abundances with a surface gravity log~{\it g} of 4.5 dex. The resolving power of the {\tt TLUSTY} spectra has been degraded to match that of {\tt ATLAS9}, which is approximately 300.}
    \label{fig:5800A}
\end{figure}

We present a comparison of spectra from {\tt ATLAS9} and {\tt TLUSTY} in the wavelength range of 5600 to 6200~\AA. As illustrated in Figure \ref{fig:5800A}, the  absorption features near 5881\AA~ in the {\tt ATLAS9} spectra exhibit abnormal profiles and show significant deviation from those in the {\tt TLUSTY} spectra when $T_{\rm eff}$ exceeds 30,000~K.
\end{document}